\documentclass[aps,prl,twocolumn,showpacs,floatfix]{revtex4}
\usepackage{graphicx}
\usepackage{times}
\usepackage{nicefrac}
\usepackage{amsmath}
\usepackage{amsfonts}
\usepackage{amssymb}
\usepackage{amsthm}
\usepackage{epsf}
\usepackage{bm}
\usepackage{bbm}
\usepackage{color}

\usepackage{dcolumn}
\newcolumntype{.}{D{x}{}{-1}}

\newcommand{\be}{\begin{eqnarray}}
\newcommand{\ee}{\end{eqnarray}}
\newcommand{\la}{\langle}
\newcommand{\ra}{\rangle}

\newcommand{\veps}{\varepsilon}

\newcommand{\bfR}{{\bf R}}

\newcommand{\bfr}{{\bf r}}

\begin{document}

\title{Nuclear excitation by two-photon electron transition}

\author{A. V. Volotka,$^{1,2}$ A. Surzhykov,$^{3,4}$ S. Trotsenko,$^{1,5}$ G. Plunien,$^{6}$ Th. St\"ohlker,$^{1,5,7}$ and S. Fritzsche$^{1,8}$}

\affiliation{
$^1$ Helmholtz-Institut Jena, D-07743 Jena, Germany\\
$^2$ Department of Physics, St. Petersburg State University, 198504 St. Petersburg, Russia\\
$^3$ Physikalisch-Technische Bundesanstalt, D-38116 Braunschweig, Germany\\
$^4$ Technische Universit\"at Braunschweig, D-38106 Braunschweig, Germany\\
$^5$ Institut f\"ur Optik und Quantenelektronik, Friedrich-Schiller-Universit\"at, D-07743 Jena, Germany\\
$^6$ Institut f\"ur Theoretische Physik, Technische Universit\"at Dresden, D-01062 Dresden, Germany\\
$^7$ GSI Helmholtzzentrum f\"ur Schwerionenforschung, D-64291 Darmstadt, Germany\\
$^8$ Theoretisch-Physikalisches Institut, Friedrich-Schiller-Universit\"at, D-07743 Jena, Germany\\
}

\begin{abstract}
A new mechanism of nuclear excitation via two-photon electron transitions (NETP) is proposed and studied theoretically.
As a generic example, detailed calculations are performed for the E1E1 $1s2s\,^1S_0 \rightarrow 1s^2\,^1S_0$ two-photon decay
of He-like $^{225}$Ac$^{87+}$ ion with the resonant excitation of the $3/2+$ nuclear state with the energy 40.09(5) keV.
The probability for such a two-photon decay via the nuclear excitation is found to be $P_{\rm NETP} = 3.5 \times 10^{-9}$
and, thus, is comparable with other mechanisms, such as nuclear excitation by electron transition and by electron capture.
The possibility for the experimental observation of the proposed mechanism is thoroughly discussed.
\end{abstract}

\pacs{31.30.J-, 32.80.Wr, 23.20.Lv, 25.20.Dc}

\maketitle
%
%
Atomic physics has kept a tenable position for many decades in the foundation and development of our knowledge
on nuclear properties. In particular, much informations about the nuclear spins, nuclear magnetic moments, and
mean-square charge radii originate from atomic spectroscopy \cite{kluge:2010:295}. Apart from the properties
of the nuclear ground or isomeric states, atomic spectroscopy provides also access to the internal nuclear
dynamics. For instance, nuclear polarization effects, that arise due to real or virtual nuclear electromagnetic
excitations, play a paramount role in an accurate description of muonic atoms \cite{borie:1982:67}. Many
years passed after they have been consistently incorporated within the framework of relativistic bound-state
QED \cite{plunien:1995:1119}. Today, the precision in determining the transition energies in highly charged
ions requires to account for the nuclear polarization corrections \cite{volotka:2014:023002}. In addition, the
single nuclear resonances can be also accessed with certain electron transitions.

The accurate determination of nuclear excitation energies and transition rates provides information not only
about the nuclear structure of individual isotopes, but also gives access to a number of gripping
applications. In the past, for example, two mechanisms were proposed for nuclear excitations by using the
techniques of atomic spectroscopy. A first one suggested by Morita \cite{morita:1973:1574} is known as nuclear
excitation by electron transition (NEET). In this process, bound-electron transitions may resonantly induce
nearly degenerate nuclear excitations. Another mechanism, the nuclear excitation by electron capture (NEEC),
was later suggested by Goldanskii and Namiot \cite{goldanskii:1976:393} and describes the resonant capture of
a free electron with the simultaneous excitation of the nucleus. In this latter case, the energy due to the
capture of the electron is transferred to nuclear internal degree of freedom and subsequently released by the
nuclear deexcitation. The scenario of the NEEC process with subsequent x-ray emission relevant for highly
charged ions was proposed in Ref.~\cite{palffy:2008:330}. However, since the nuclear resonances are very narrow,
for both mechanisms, NEET and NEEC, it is extremely important to finely adjust the atomic and nuclear transition
energies to each other, and this makes the observations of these processes rather challenging. Indeed, only the
NEET process has so far been verified experimentally for $^{197}$Au \cite{kishimoto:2000:1831,kishimoto:2006:031301},
$^{189}$Os \cite{aoki:2001:044609}, and $^{193}$Ir \cite{kishimoto:2005:3} atoms.

Further studies of the nuclear excitation mechanisms by atomic transition enable us not only to better
understand the interactions between the nucleus and electrons and to determine nuclear parameters, but also
opens perspectives to a variety of fascinating applications. One among them is the access to low-lying
isomeric nuclear excitations, e.g., the isomeric states $^{229m}$Th \cite{inamura:2009:034313,
raeder:2011:165005,wense:2016:47} and $^{235m}$U \cite{chodash:2016:034610} with an excitation energy of
several (tens) eV. Other potential applications can be seen in the isotope separation \cite{morita:1973:1574},
energy storage \cite{walker:1999:35} and its controlled release \cite{tkalya:2005:525,palffy:2007:172502}.

%
\begin{figure*}
\includegraphics[width=0.95\textwidth]{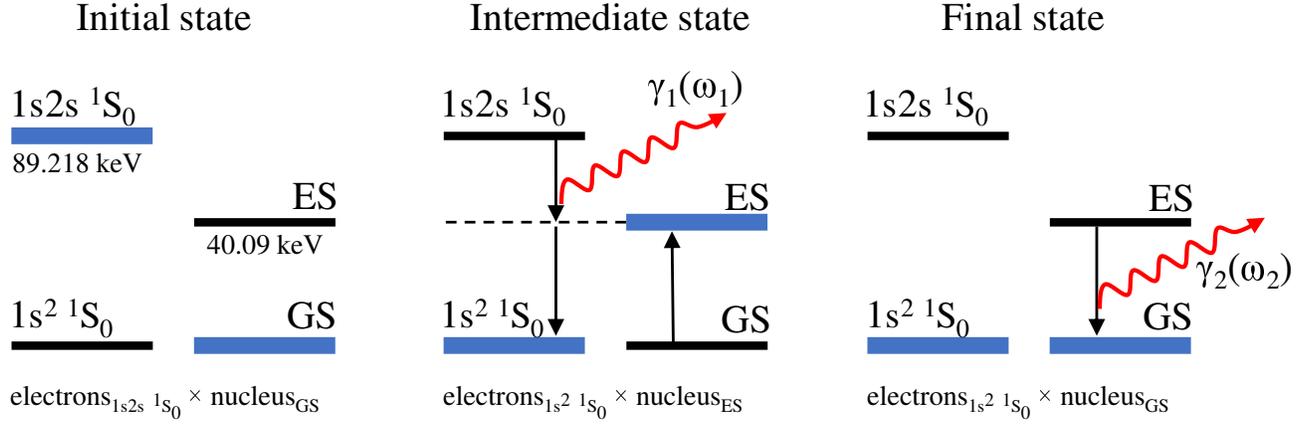}
\caption{(Color online) The mechanism of the nuclear excitation by two-photon electron transition as here
explained for He-like $^{225}$Ac$^{87+}$ ion. The initial state (left panel), which is characterized by the
$1s2s\,^1S_0$ electronic state and nuclear ground state (GS), decays into the final state (right panel), where
both the electrons and nucleus are in their ground states $1s^2\,^1S_0$ and GS, respectively, via the
intermediate cascade state (middle panel) with the nucleus being in the excited state (ES). The emitted
photons $\gamma_1$ (electron decay photon) and $\gamma_2$ (nuclear fluorescence photon) are depicted by wavy
lines with arrows.
\label{fig:1}}
\end{figure*}
In this Letter, we present and discuss a new mechanism for nuclear excitation to which we refer as nuclear
excitation by two-photon electron transition (NETP). An electron transition can proceed via emission of not
only one photon, but also via simultaneous emission of two photons, which share the transition energy. In
contrast to the one-photon transitions, where the photon frequency equals the transition energy, the energy
distribution of the spontaneously emitted photons then forms a continuous spectrum. This implies, that some
of the photons exactly match in their frequency with the nuclear transition energy as long as the nuclear
excitation energy is smaller than the total electron transition energy. In this way, a nucleus resonantly
absorbs this photon and gets excited. This mechanism can also be understood as the two-photon electron
transition in the presence of intermediate (nuclear) cascade states. In the case of NETP, the electrons and
the nucleus must be treated as combined system in which the intermediate cascade state is given by the excited
nucleus and the electrons in their ground level. Similarly as for a pure electronic two-photon decay, the
presence of a cascade essentially increases the photon emission intensity in the region of the resonant energy.
A key advantage of the NETP process is that, in contrast to the NEET and NEEC, such resonant nuclear
excitations may happen for all nuclear levels with an access energy smaller than the total transition energy.
In the following, we derive the formulas describing the NETP mechanism and perform calculations especially
for the two-photon decay $1s2s\,^1S_0 \rightarrow 1s^2\,^1S_0$ in He-like $^{225}$Ac$^{87+}$ ion. We find that
the probability of the two-photon decay via nuclear excitation is surprisingly large
$P_{\rm NETP} = 3.5 \times 10^{-9}$ and comparable with the corresponding NEET probability values
$P_{\rm NEET}$ of previously observed \cite{kishimoto:2000:1831,kishimoto:2006:031301,aoki:2001:044609,
kishimoto:2005:3} as well as theoretically proposed scenarios \cite{tkalya:1992:209,harston:2001:447}.

The NETP process is shown as a two-step process in Fig.~\ref{fig:1} in a more picturesque way. For the sake of
clarity and without losing generality, we shall refer below always to He-like $^{225}$Ac$^{87+}$ ion. In
the initial state the electrons are in the excited state $1s2s\,^1S_0$ and the nucleus is in its ground state
(GS). Then, the electrons undergo the two-photon decay into its ground state $1s^2\,^1S_0$ via the
intermediate state and the electron decay photon $\gamma_1$ with the energy $\omega_1$ is emitted. In the
second step, the nucleus being in the excited state (ES) radiatively decays into its GS with an emission of
the nuclear fluorescence photon $\gamma_2$ with the energy $\omega_2$. Due to energy conservation, the sum of
the photon energies is equal to the total energy $\Delta E$ of the electron transition $1s2s\,^1S_0
\rightarrow 1s^2\,^1S_0$, i.e., $\Delta E = \omega_1 + \omega_2$. The E1E1 two-photon transition $1s2s\,^1S_0
\rightarrow 1s^2\,^1S_0$ in $^{225}$Ac$^{87+}$ ion is chosen here for various reasons. For such ions, first,
the two-photon transition happens rather fast with the total rate $W_{1s2s\,^1S_0} = 6.002 \times 10^{12}$
s$^{-1}$ \cite{volotka:2011:062508} and defines the lifetime $\tau_{1s2s\,^1S_0} = 0.167$ ps of the
$1s2s\,^1S_0$ level completely. Second, the $1s2s\,^1S_0$ state can be produced quite selectively in
collisions of Li-like ions with gas atoms \cite{fritzsche:2005:S707,rzadkiewicz:2006:012511} and,
moreover, the two-photon decay energy spectrum has been accurately measured for He-like Sn$^{48+}$
\cite{trotsenko:2010:033001} and U$^{90+}$ \cite{banas:2013:062510} ions. For $^{225}$Ac$^{87+}$ ion, the
emitted photons span the frequency region up to the total transition energy $\Delta E = 89.218(2)$ keV
\cite{artemyev:2005:062104}. As for the probing nuclear excitation resonance, which lies inside the spanned
energy region, we take the $3/2+$ level of $^{225}$Ac nucleus with the excitation energy $\omega_{\rm ES} =
40.09(5)$ keV \cite{225:2009}. This ES in the case of neutral actinium atom has a half-lifetime 0.72(3) ns
and decays primarily into the GS via the electric-dipole photon or conversion electron emission with a total
conversion coefficient of $\simeq 1$ \cite{ishii:1985:237}. For He-like $^{225}$Ac$^{87+}$ ion, we, therefore,
need to consider only the radiative E1 deexcitation channel with the transition rate $W_{\rm ES} = 0.41 \times
10^9$ s$^{-1}$ and the corresponding linewidth $\Gamma_{\rm ES} = 2.7 \times 10^{-7}$ eV.

\begin{figure}
\includegraphics[scale=0.7]{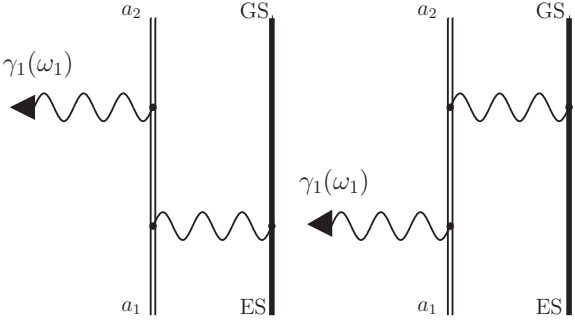}
\caption{Feynman diagrams that represent the nuclear excitation during the two-photon electron transition from
an initial state $a_2$ to a final state $a_1$. The double lines indicate the electron wave functions and
electron propagator in the Coulomb field of the nucleus, the heavy lines denote the nucleus in its ground (GS)
and excited (ES) states, and the internal wavy line stays for the photon propagator. The emission of the
electron decay photon $\gamma_1$ is depicted by the wavy line with outgoing arrow.
\label{fig:1a}}
\end{figure}

Now let us provide the theoretical formalism describing the NETP mechanism. While the second-step process is
fully determined by the nuclear decay rate itself $W_{\rm ES}$, the description of the first step, i.e., the
nuclear excitation, has to be formulated. Fig.~\ref{fig:1a} displays the Feynman diagrams that describe the
first-step process.
The corresponding $S$-matrix element is of third order and can be written (in relativistic units $\hbar = 1,\,
c = 1,\,m = 1$) by following the basic principles of QED \cite{berestetsky}:
\allowdisplaybreaks
\be
\label{eq:1}
S^{(3)}_{\rm NETP} &=& \frac{1}{\Delta E - \omega_{\rm ES} - \omega_1 - i(\Gamma_{1s2s\,^1S_0}+\Gamma_{\rm ES})/2}
                       \nonumber\\
              &\times& \frac{e^2}{4\pi}
                       \int d^3r_1 d^3r_2 d^3R\;
                       \overline{\psi}_{a_1}(\bfr_1)
                       \nonumber\\
              &\times& \left\{
\gamma_0\frac{1}{|\bfr_1-\bfR|} S(\veps_{a_2}-\omega_1,\bfr_1,\bfr_2) \gamma^\mu A^*_\mu(\omega_1,\bfr_2)
               \right. \nonumber\\
                   &+& \left.
\gamma^\mu A^*_\mu(\omega_1,\bfr_1) S(\veps_{a_1}+\omega_1,\bfr_1,\bfr_2) \gamma_0\frac{1}{|\bfr_2-\bfR|}
                       \right\}
                       \nonumber\\
              &\times& {\psi}_{a_2}(\bfr_2)\,
                       \Psi^\dagger_{\rm ES}(\bfR) \hat{\rho}_{\rm fluc}(\bfR) \Psi_{\rm GS}(\bfR) \,,
\ee
where $\bfr_1$ and $\bfr_2$ are electron coordinates, and $\bfR$ is nuclear coordinate. Moreover,
$\Gamma_{1s2s\,^1S_0}$ and $\Gamma_{\rm ES}$ denote the widths of the $1s2s\,^1S_0$ electronic level and the
nuclear excited state, while the electron wave functions $\psi_{a_1}$ and $\psi_{a_2}$ are bound-state
solutions of the Dirac equation for the $1s$ and $2s$ states, respectively. The wave functions $\Psi_{\rm GS}$
and $\Psi_{\rm ES}$ describe the nucleus in its ground and excited states. $\gamma^\mu$ are the Dirac
matrices, $S(\omega,\bfr_1,\bfr_2)$ is the electron propagator, and $A^*_\mu(\omega,\bfr)$ is the emitted
photon wave function. The electron-nucleus interaction acts via the photon propagator and which is taken in
Coulomb gauge and just restricted to the Coulomb term only. The nuclear charge-density operator
$\hat{\rho}_{\rm fluc}$ characterizes the intrinsic nuclear dynamics due to external electromagnetic
excitations and could be decomposed in terms of nuclear multipoles as discussed in
Refs.~\cite{plunien:1991:5853,plunien:1995:1119}. Eq.~(\ref{eq:1}) was obtained in the resonant approximation,
i.e., $\omega_1 \approx \Delta E - \omega_{\rm ES}$, and after integration over the time variables in all three
vertexes. It should be mentioned here, that here we neglect the interference term between NETP and {\it pure}
two-photon electron transition, since it turns out to be negligible small in the present scenario. We finally
note also, that the expression obtained for the $S$-matrix element is quite general and applies similarly for
any other NETP scenario.

To evaluate the $S$-matrix element in Eq.~(\ref{eq:1}), we follow the standard procedures. Making use of the
multipole expansion of the (Coulomb-) photon propagator, we can factorize the nuclear variables and arrive
immediately at the matrix element of the nuclear electric transition operator $\hat{Q}_{LM}$:
\be
\label{eq:2}
\la I_{\rm ES} M_{\rm ES} |\hat{Q}_{LM}| I_{\rm GS} M_{\rm GS} \ra &=&
           \int d^3R\;
           \Psi^\dagger_{\rm ES}(\bfR) \hat{\rho}_{\rm fluc}(\bfR)\nonumber\\
  &\times& \Psi_{\rm GS}(\bfR)\,
           R^L Y^*_{LM}(\hat{\bfR})\,,
\ee
where $I_{\rm ES}$, $M_{\rm ES}$ and $I_{\rm GS}$, $M_{\rm GS}$ are the nuclear spins and their (magnetic)
projections for the excited and ground nuclear states, respectively. Then, the square of the reduced matrix
element of the transition operator $\hat{Q}_{LM}$ can be commonly expressed in terms of the reduced transition
probability $B(EL;I_{\rm GS} \rightarrow I_{\rm ES})$. We note here, that in accordance with the multipole
expansion the nuclear excitation must have the same type (magnetic or electric) and multipolarity as the
one-electron transition, which it replaces in the normal two-photon transition amplitude. If, however, the
nuclear and electronic variables are disentangled, we can employ experimental data for the reduced transition
probability \cite{225:2009}. The remaining electronic part in the $S$-matrix element is evaluated here
similarly as in Ref.~\cite{volotka:2014:023002}. The dual-kinetic-balance finite basis set method
\cite{shabaev:2004:130405} is employed to represent the Dirac spectrum in the Coulomb potential of an extended
nucleus. Knowing the $S$-matrix element one can easily obtain the total rate of the NETP process $W_{\rm NETP}$
as square of the modulus of the $S$-matrix element integrated over the energy of the emitted photon $\omega_1$
and multiplied by the total width of the process $\Gamma_{1s2s\,^1S_0} + \Gamma_{\rm ES}$. As a result, we find
the rate $W_{\rm NETP} = 0.21 \times 10^5$ s$^{-1}$ for He-like $^{225}$Ac$^{87+}$ ion. Furthermore, in order
to compare NETP and two-photon probabilities, we define the dimensionless ``NETP probability''
$P_{\rm NETP} = W_{\rm NETP} / W_{1s2s\,^1S_0}$, which determines the (relative) probability that the decay of
the initial atomic state $1s2s\,^1S_0$ will proceed via the excitation of the nucleus. For the given example,
we here receive $P_{\rm NETP} = 3.5 \times 10^{-9}$ and, thus, a relative rate that this comparable with the
corresponding values for the NEET process, $P_{\rm NEET} \propto 10^{-7} ... 10^{-12}$, for most of proposed
examples \cite{tkalya:1992:209,harston:2001:447}. When the nucleus got excited by the NETP process
(cf. Fig.~\ref{fig:1}) it decays to the nuclear GS with the transition rate $W_{\rm ES}$, the linewidth
$\Gamma_{\rm ES}$ and under the emission of a nuclear fluorescence photon $\gamma_2$ with energy
$\omega_2 = \omega_{\rm ES}$.

\begin{figure}
\includegraphics[scale=0.35]{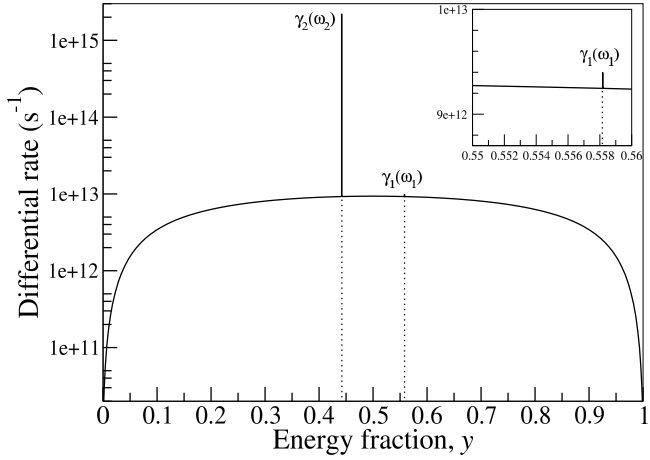}
\caption{$1s2s\,^1S_0 \rightarrow 1s^2\,^1S_0$ two-photon differential rate plotted as a function of the
reduced energy (sharing fraction) $y$ for He-like $^{225}$Ac$^{87+}$ ion. The NETP resonances are
emphasized by the dotted lines.
\label{fig:2}}
\end{figure}

Now let us discuss the possibility of the experimental observation of the NETP mechanism. The presence of
additional decay channel significantly modifies the energy spectrum of the usual two-photon emission in the
vicinity of the nuclear resonance energy. In Fig.~\ref{fig:2} the energy-differential rate for the decay of
the $1s2s\,^1S_0$ state is displayed as a function of the reduced energy $y = \omega / \Delta E$, where $\omega$
is the energy carried by one of the emitted photons. As one can see from the figure, the NETP mechanism leads
to the appearance of two peaks: the first one at the energy $\omega \approx \omega_1$ and with the width
$\Gamma_{\rm ES}$, while the second one at $\omega \approx \omega_2$ has the width  $\Gamma_{1s2s\,^1S_0} +
\Gamma_{\rm ES}$. Due to these features of the expected energy sharing of the emitted photons, one can think
of two possible options for the experimental observation of the NETP process, which consist in the
measurements of either the electron decay $\gamma_1$ or nuclear fluorescence $\gamma_2$ photons, respectively.
If we first consider the observation of photons with frequency $\omega_1$, the emission of the $\gamma_1$
photon cannot be separated (in time) from the background signal that is formed by the {\it pure} two-photon
electronic decay, since both just follow the population of the $1s2s\,^1S_0$ state. Therefore, the fluorescence
intensity of $\gamma_1$ photons $I_{\gamma_1}(t)$ decays within the same time, $I_{\gamma_1}(t) \sim
{\rm exp}(-t\,W_{1s2s\,^1S_0})$ and, hence, the main difficulty is to resolve the NETP photons from the
background. The signal-to-background ratio can be determined by the partial NETP probability $p_{\rm NETP}(\Delta)$,
which is defined as
\be
p_{\rm NETP}(\Delta) = \frac{W_{\rm NETP}}
                            {\displaystyle\int_{\omega_1 + \Delta/2}^{\omega_1 - \Delta/2}
                             dW_{1s2s\,^1S_0}(\omega)}
\ee
and, which describes the probability that a photon with an energy in the range between $\omega_1 - \Delta/2$
and $\omega_1 + \Delta/2$ is emitted via the NETP process. Here, $dW_{1s2s\,^1S_0}(\omega)$ is the
energy-differential rate of the electron two-photon transition and $\Delta$ corresponds to the energy interval
that can be distinguished experimentally. For typical x-ray detectors with a resolutions of, say, $\Delta = 1$
eV, $10$ eV, or $100$ eV, we, therefore, get $p_{\rm NETP}(1\,{\rm eV}) = 1 \times 10^{-4}$,
$p_{\rm NETP}(10\,{\rm eV}) = 1 \times 10^{-5}$, or $p_{\rm NETP}(100\,{\rm eV}) = 1 \times 10^{-6}$,
respectively. Recent progress in developing x-rays detectors enabled one to drastically increase their
resolution up to the level of 5 eV or even better, and with a gain in efficiency,
cf.~Ref.~\cite{hengstler:2015:014054}. In this regard, the separation of $\gamma_1$ photons might be achieved
soon already with present or near-future x-ray technology.

\begin{figure}
\includegraphics[scale=0.35]{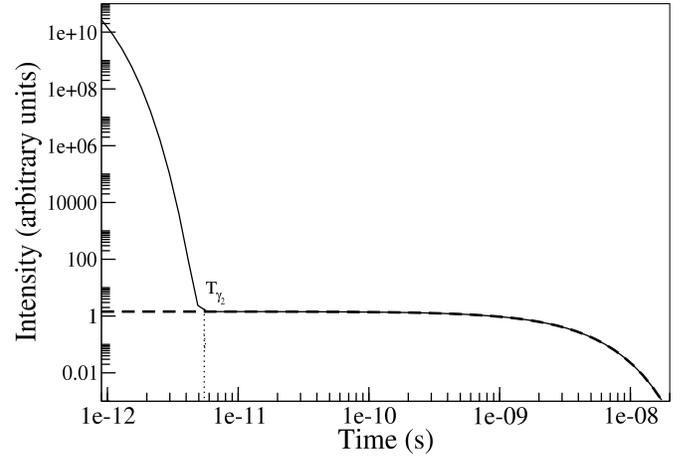}
\caption{The overall intensity produced by the decay of the $1s2s\,^1S_0$ level (solid line) is compared with the
intensity of nuclear fluorescence $\gamma_2$ photons (dashed line), which are plotted as functions of time in
arbitrary units.
\label{fig:3}}
\end{figure}

A second set-up of experiments refers to the observation of the nuclear fluorescence $\gamma_2$ photons. In
contrast to an enhanced emission of $\gamma_1$ photons, the $\gamma_2$ fluorescence occurs with a certain time
delay, which corresponds to the difference between the lifetimes of the $1s2s\,^1S_0$ state (0.167 ps) and the
nuclear excited state (2.4 ns). We can express the intensity of this $\gamma_2$ fluorescence as function of time,
\be
I_{\gamma_2}(t) \sim {\rm exp}(-t\,W_{\rm ES}) - {\rm exp}(-t\,W_{1s2s\,^1S_0})\,,
\ee
and display it in Fig.~\ref{fig:3} together with the overall and continuous photon intensity due to the decay
of the $1s2s\,^1S_0$ state (NETP and the {\it pure} two-photon decay). As seen from this figure, one can clearly
identify the emission of $\gamma_2$ photons by observing the fluorescence after some small time delay of, say,
$T_{\gamma_2} = 5$ ps, at which the background intensity from the electronic two-photon decay will already be
strongly reduced. If we now define the time-dependent NETP probability $p_{\rm NETP}(T)$ as a relative
probability that the photon emitted at time $T$ originates from the NETP process, for times larger than
$T_{\gamma_2}$ it tends to one, i.e., $p_{\rm NETP}(T > T_{\gamma_2}) \approx 1$. Thus, the observation of
$\gamma_2$ photons emission actually serves us as signature of the NETP process. In this regard, the
measurement of $\gamma_2$ photons seems to be presently more feasible for verifying the NETP process.

The latter scenario is planned to be realized at the current GSI (Darmstadt) facility \cite{trotsenko:2016:LI}.
The initial $1s2s\,^1S_0$ state can be efficiently produced in the collision of Li-like ions of the given
isotope with N$_2$ gas target via the selective $K$-shell ionization \cite{rzadkiewicz:2006:012511}. Since the
$1s2s\,^1S_0$ state almost exclusively decays via the two-photon transition into the ground state, the most of
produced He-like ions contribute to the process under consideration. The x-ray emission will be measured in
time-coincidences with the detection of the up-charged (He-like) ions, whose efficiency is almost 100\%. All
these will enable us to measure a very clean spectrum of the two-photon decay \cite{trotsenko:2010:033001,
banas:2013:062510}. In order to observe the delayed nuclear fluorescence photons $\gamma_2$, a high-efficiency
in-vacuum x-ray detector will be installed to cover a solid angle as large as possible. Fast transitions, that
will mostly decay in the vicinity of the gas-target, will be shielded in order to reduce background in the
measurement of the delayed photons. In the first experiment, we will compare the measured intensity for
$^{225}$Ac and another isotope (e.g., $^{227}$Ac) ion beams in order to unambiguously verify the delayed emission
of $40$ keV photons. Later measurements will record the x-ray intensities at different distances from the
gas-target, which in turn will allow us to measure the NETP probability $P_{\rm NETP}$ in a way similar to the
beam-foil spectroscopy technique \cite{traebert:2008:038103}. At the experimental storage ring (ESR) at GSI beams
of $\gtrsim 10^{8}$ cooled ions can be provided and stored for collisions with the gas-jet target with the areal
densities above $10^{14}$ cm$^{-2}$ \cite{petridis:2014:TR,petridis:2015:014051}. Because of the high revolution
frequencies of ions in the storage ring (about 2 MHz) and the recurring interaction of ions and target electrons,
a very high luminosity can be achieved. Ultimately, we expect stimulating of up to few hundreds NETP fluorescence
photons per day of the beamtime. This looks very feasible for the successful observation and characterization of
the NETP process. Moreover, at the new FAIR accelerator complex the experiment will profit from the higher
luminosity as well as from the ability of the measurements much closer to the ion beam.

%
In conclusion, we here present a new mechanism for nuclear excitation by two-photon electron transition (NETP).
In contrast to the previously suggested mechanisms, NEET and NEEC, there is no need for observing this
mechanism to adjust the electronic and nuclear transition energies to each other. Instead, we can simply
utilize the continuous spectrum of the two-photon decay in order to scan for the appropriate nuclear excitation
levels. For the given example of the E1E1 two-photon transition $1s2s\,^1S_0 \rightarrow 1s^2\,^1S_0$ in
He-like $^{225}$Ac$^{87+}$ ion, we predict the probability $P_{\rm NETP} = 3.5 \times 10^{-9}$ when compared
with the overall and continuous two-photon emission.

Apart from probing our understanding of the electron-nucleus interaction and nuclear structure, the experimental
verification of the NETP process may have far reaching consequences, such as for the search of low-lying isomeric
states, for energy storage and its release in a controlled manner \cite{walker:1999:35,tkalya:2005:525,
palffy:2007:172502}, or elsewhere. We, therefore, hope that this work lays the foundation for developing NETP
processes as a sensitive tool at the borderline of atomic and nuclear physics.

%
This work was supported by BMBF Project No. 05P15SJFAA.


%
\end{document}